\documentclass[nohyper,notoc]{JHEP}
\usepackage{epsfig,axodraw}
\def\eq{{\rm eq}}
\def\ann{{\rm ann}}
\def\c{{\rm c}}
\def\elas{{\rm elas}}
\def\therm{{\rm therm}}
\def\GeV{{\rm GeV}}
\def\GUT{{\rm GUT}}
\def\inel{{\rm inel}}
\def\in{{\rm in}}
\def\max{{\rm max}}
\def\min{{\rm min}}
\def\Pl{{\rm Pl}}
\def\rad{{\rm rad}}
\def\reh{{\rm reh}}
\def\rmd{{\rm d}}
\def\rme{{\rm e}}
\def\scat{{\rm scat}}
\def\vev#1{{\langle#1\rangle}}
\def\lesssim{\mathrel{\vcenter
     {\hbox{$<$}\nointerlineskip\hbox{$\sim$}}}}
\def \gtrsim{\mathrel{\vcenter
     {\hbox{$>$}\nointerlineskip\hbox{$\sim$}}}}

\preprint{OUTP-00-38P}
\title{Thermalisation after inflation}

\author{ Sacha Davidson and Subir Sarkar
 \\Theoretical Physics, University of Oxford, 1 Keble Road,
Oxford OX1 3NP, UK }
\abstract{
During (re)heating of the Universe after 
inflation, the relativistic
decay products $\chi$ of the inflaton  must lose energy, and
additional particles must be produced, to attain a thermalised state at
a temperature $T_{\reh}$. We estimate the rate of energy loss via 
elastic and inelastic scattering interactions. Elastic scattering is
an inefficient energy loss mechanism so inelastic processes, although
higher order in the coupling $\alpha$, can be faster because more
energy is transfered. The timescale to produce a particle number
density of ${\cal O}(T_{\reh}^3)$ is the inelastic energy loss
timescale $\sim(\alpha^3 T_{\reh}^2/E_{\chi})^{-1}$. 
The Universe will be thermalised within a Hubble time
at $T_\reh$ if $E_{\chi} \lesssim \alpha^3 M_{Pl}$.}

\begin{document}

\section{Introduction}
\label{intro}

The primordial density fluctuations required to account for the
observed large-scale structure in the universe can plausibly be
generated during a quasi De Sitter expansion phase at early times
\cite{lss}. The recent detection \cite{cmb} of the expected
characteristic signature of such fluctuations in the cosmic microwave
background anisotropy provides strong evidence in favour of such an
inflationary phase during which the Universe underwent superluminal
expansion while dominated by the vacuum energy of a scalar field ---
the inflaton \cite{infl}. This period must last long enough to
generate a homogeneous universe with small density fluctuations up to
the scale of at least the present Hubble radius, but must eventually
evolve into the hot radiation dominated era of the standard Big Bang
model. The process by which the vacuum energy is converted into
relativistic particles can be quite complex. Traditionally only the
perturbative decay of the inflaton \cite{AFW}
 has been considered for studying
(re)heating, but there may also be non-perturbative
transfer of energy to other fields --- dubbed ``preheating''
\cite{preheat}. However, this is not always an efficient energy
loss mechanism for the inflaton when its decay products
have self-interactions \cite{pr,igor}. Field theoretical studies of the 
energy transfer from a cosmological scalar field to other particles have 
been performed, but so far only in the context of toy models \cite{bvhs}. Our
interest here is to determine the time-scale for the thermalisation of the
bulk of the vacuum energy,
when the inflaton can be taken to decay {\em perturbatively} into
relativistic particles.

The momentum space distribution of the relativistic decay products
($\chi$) of the inflaton ($\phi$) will not initially be thermal, so
these particles must interact with each other to redistribute their
momenta as well as 
 adjust the particle number density
to create
a thermal distribution. 
This process 
is called 
``thermalisation'', and has been discussed
previously by many authors \cite{EENO,scott,enq,enq2,ZPM,McT,R}.
There are
two aspects to thermalisation --- reaching {\em kinetic} equilibrium,
and achieving {\em chemical} equilibrium. For the first, the momentum
must be redistributed among the particles present, which can happen
via $2\to2$ scatterings and annihilations. For the second, the
comoving particle number density must be modified, e.g. by decays,
inverse decays,  or
$2\to3$ particle interactions.
 We assume here that the inflaton
decay products are initially
more energetic and less dense than the distribution
to which they thermalise. This means
that  the
inflaton decay products must lose energy and produce additional
particles. This corresponds to generic 
``new''  inflationary
models, but
possibly not to ``chaotic" models where inflation occurs at field
values beyond the Planck scale.

It has been noted \cite{EENO,scott} that the inflaton decay products might
not be thermalised at $T_{\reh}$, defined as the temperature when they
first dominate the energy density of the Universe. Thermalisation was
estimated  in \cite{EENO} to occur when the $\chi\bar{\chi}$ annihilation rate
$\Gamma_{\ann}$ begins to exceed the Hubble expansion rate $H$, which
happens well below $T_{\reh}$ in many inflation models. A more detailed
analysis using the Boltzman equation can be found in \cite{scott}.   
In a numerical
study of the thermalisation of a gas of semi-classical particles
\cite{enq} it was found that the particles indeed reach kinetic
equilibrium after a few hard scattering interactions, i.e. when
$\Gamma_{\ann}\sim\,H$. However it takes longer to achieve chemical
equlibrium since this requires new particle production; it was argued
in \cite{enq} that the timescale for this is of order $\alpha^{-1}$
times the kinetic equilibration timescale.

It has been suggested that soft processes can make thermalisation
faster \cite{ckr,McT}, because the interaction rate for processes with
small momentum transfer is larger than the hard scattering rate used
in ref.\cite{EENO,scott}. It has also been argued that low energy particles
can act as a seed for thermalising the energetic inflaton decay
products, because the cross-section for annihilation with a soft
particle is larger than with an energetic particle \cite{McT}. This
scenario of catalysed thermalisation could be relevant if particles
reach a thermal distribution via $2\to2$ interactions and decays.
Thermalisation via annihilation and decays has been recently
discussed, in a Universe where the baryon asymmetry is generated via
the Affleck-Dine mechanism \cite{R}. This is complementary to the
present work, where we will concentrate on thermalisation via $2\to2$
and $2\to3$ scattering interactions.

In this paper we discuss which interactions will thermalise a bath of
relativistic fermions with gauge interactions produced in inflaton
decay. The thermalisation timescale is usually taken to be an
interaction timescale, which begs the question ``which interaction?''.
Thermalisation can be very fast, if it happens on the timescale of
soft scattering processes, because these cross-sections diverge as the
momentum transfer goes to zero. At the opposite extreme, the timescale
can exceed the Hubble time, if thermalisation requires hard processes
with momentum transfer of order the incident particle energy. We argue
that the thermalisation timescale is at least as long as the timescale
for an inflaton decay product, produced with energy of ${\cal
O}(m_\phi)$, to lose an energy $\sim(m_{\phi}-T_{\reh})$. We estimate
$\Gamma_{\elas}\,(=\rmd(\ln{E})/{\rmd}t)$, the rate of energy loss via
{\em elastic scattering} of an energetic particle incident on other
energetic particles, and find that the timescale for it to lose its
incident energy is the hard annihilation timescale. We estimate the
rate for energy loss due to {\em inelastic} $2\to3$ scattering, and
find $\Gamma_{\inel}\gg\Gamma_\elas$. This suggests that $2\to3$
scattering interactions can thermalise the Universe faster than
$2\to2$ processes and decays, because although higher order in
$\alpha$, $2\to3$ interactions are lower order in
$T_{\reh}/m_\phi$. Such processes are also neccessary to bring the
particles into chemical equilibrium. The timescale to produce a
particle number density of ${\cal O}(T_{\reh}^3)$ via $2\to3$
interactions is given by $\Gamma_{\inel}^{-1}$, which is therefore the
true thermalisation timescale.

In the next section we introduce our model and review relevant
previous work \cite{enq,McT,R}. Its purpose is to introduce notation
and make this paper self-contained. In the following two sections, we
discuss what interaction should be used to estimate the thermalisation
rate. We study the rate of energy loss of a particle via elastic
scattering in section~\ref{sect3}. This process is solvable and only
has a logarithmic infrared divergence. In section~\ref{sect4}, we
consider inelastic $2\to3$ processes. We estimate the timescale for an
inflaton decay product to lose an energy $\sim\,m_\phi$, and the
timescale to produce a number density $\sim\,T_{\reh}^3$. In section
\ref{sect5}, we outline the thermalisation discussion of previous
papers, which uses the annihilation rate as the thermalisation
rate. We then repeat the analysis using the inelastic rate, which
suggests that the Universe will indeed be thermal at $T_{\reh}$. We
present our conclusions in section~\ref{sect6}.

\section{Model}
\label{sect2}

We consider a scalar field $\phi$ whose potential energy is the
principle component of the energy density of the Universe $\rho$. (We
refer to $\phi$ as the inflaton, but it could be any other scalar
field, e.g. a modulus, which dominates the universe.) The Hubble
expansion rate when $\phi$ starts oscillating coherently about the
minimum of its potential is 
\begin{equation}
 H_{\in}^2 \simeq \frac{8\pi}{3} \frac{\rho_\phi(a_{\in})}{M_{\rm Pl}^2} 
   \simeq m_{\phi}^2.  \end{equation} The field $\phi$ decays at a
rate $\Gamma_\phi\equiv\alpha_\phi\,m_\phi$ to two light fermions
$\chi$ and $\bar{\chi}$. \footnote{Our results are largely based on
dimensional analysis, so we would not expect them to change if the
inflaton decay products were different.} The energy density of
coherent scalar field oscillations redshifts like matter as the
Universe expands, so the $\phi$ energy density $\rho_\phi$ then
decreases with time $\tau$ as $a^{-3}{\rme}^{-\Gamma\tau}$. The
Universe will be dominated by the $\phi$ oscillations until
$\tau\sim\Gamma_\phi^{-1}$, when most of the $\phi$ energy is
transfered to the relativistic decay products.

The ``reheat temperature'' $T_{\reh}$ is defined when
$H\sim\Gamma_\phi$ as
\begin{equation}
 \rho_{\rad} (a_{\reh}) \equiv \frac{g_* \pi^2 T_{\reh}^4}{30},
\end{equation} 
where $g_*$ is the number of relativistic degrees of freedom. The
$\chi$ particles are relativistic so we can define a ``temperature''
$T$ for this radiation, as done above. If the $\chi$ particles have
reached kinetic and chemical equilibrium, $T$ will correspond to the
thermodynamic temperature.

It is recognised \cite{infl} that there is a bath of relativistic
particles prior to $T_{\reh}$ since $\phi$ decays over time and not
instantaneously at $T_{\reh}$. There could be interesting
implications for baryogenesis and other particle abundances
\cite{Mcref,Mc,sma,gkr} if this bath is thermalised. The number
density $n_{\chi}$ of $\chi$ and $\bar\chi$ particles produced as
$\phi$ decays is
\begin{equation}
 n_{\chi} = 2 n_\phi(a_{\in}) 
 (1 - {\rme}^{-\Gamma(\tau-\tau_{\in})})
 \left(\frac{a_{\in}}{a}\right)^{3}, 
\label{nchi}
\end{equation}
where $n_\phi(a_{\in})=\rho_\phi(a_{\in})/m_\phi$. (This  neglects
$\chi$s pair-produced in $\chi$ self-interactions.)  The number density
of $\chi$s increases rapidly until $\tau \simeq 2\tau_{\in}$, then 
decreases as $a^{-3/2}$ until $\tau\sim\Gamma_\phi^{-1}$, when most of the
inflaton energy has been transfered to the $\chi$s.

After the $\chi$s are produced, their energy redshifts. A $\chi$
produced at time $\tau_1$ with energy $m_\phi/2$ will have energy
$E_2=(m_\phi/2)(a_1/a_2)=(m_\phi/2)(\tau_1/\tau_2)^{2/3}$
at time $\tau_2$, so the {\it comoving} distribution
in energy space is
\begin{equation}
 \frac{{\rmd} N}{{\rmd}E_2} = \frac{{\rmd} N}{\rmd\tau_1} \frac{\rmd\tau_1}{\rmd E_2}
 = 6\sqrt{\frac{2 E_2}{m_\phi}} 
   \frac{\Gamma\tau_2}{m_\phi} N_\phi(\tau_1) 
   \qquad (\tau_2 < \Gamma_{\phi}^{-1}),
\end{equation}
where $N(\tau)=n(\tau)(a(\tau)/a_{\in})^3$. For
$\tau_1\lesssim\Gamma^{-1}$, we can use
$N_\phi(\tau_1)\simeq\,N_\phi(a_{\in})$.
The energy density in $\chi$s at some time $\tau<\Gamma_{\phi}^{-1}$
will be
\begin{equation}
 \rho_{\chi}(a) = \int^{m_{\phi}/2} {\rmd}E E \frac{{\rmd}n}{{\rmd}E}
 \simeq \frac{3}{5} \Gamma_{\phi} (\tau - \tau_{in}) \rho_\phi(a_{\in}) 
 \left(\frac{a_{\in}}{a}\right)^3 
 \equiv \frac{g_*\pi^2}{30}{T}^4 .
\label{radT}
\end{equation} 
The maximum $\chi$ energy density (which
occurs at $\tau \simeq 2\tau_{\in}$) is 
defined  to be $g_*\pi^2T_{\max}^4/30$.
It is easy to see that  between $T_{\max}$ and $T_{\reh}$
$\rho_{\chi}$ $ \sim\,a^{-3/2}$, $T\sim\,a^{-3/8},$ and 
$(T_{\max}/T_{\reh})^{4}$ $ \simeq \alpha_{\phi}$. 

We assume that the $\chi$s have $SU(N_{\c})$ gauge interactions among
themselves with coupling $\alpha\sim1/30$. We would like to know how
soon the $\chi$ distribution will have the equilibrium form
$f(k)\sim({\rme}^{E/T}+1)^{-1}$. We can get a qualitative answer by
comparing the expansion rate $H$ to interaction rates, in which we
make some attempt to include factors of $\pi$ and $N_{\c}$ in
sections ~\ref{sect2} and \ref{sect3}. We drop them in sections
\ref{sect4} and \ref{sect5}, where the discussion is more
approximate. A more accurate result could be obtained by solving
Boltzman equations, or perhaps other more appropriate equations
\cite{Buch}, for the particle phase space distributions.

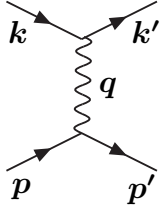
\begin{figure}[h]
\unitlength.5mm
\SetScale{1.418}
\begin{boldmath}
\begin{center}
\begin{picture}(40,60)(-20,0)
\ArrowLine(-20,5)(0,15)
\Text(-16,0)[c]{$p$}
\Text(16,0)[c]{$p'$}
\ArrowLine(0,15)(20,5)
\Photon(0.6,40.6)(0,15){2}{5}
\Text(7,27)[c]{$q$}
\ArrowLine(0,40)(20,50)
\ArrowLine(-20,50)(0,40)
\Text(-17,42)[c]{$k$}
\Text(17,42)[c]{$k'$}
\end{picture}
\end{center}
\end{boldmath}
\vspace{10mm}
\caption{$\chi $ scattering. Time runs from left
to right}
\label{f2}
\end{figure}

There are various $2\to2$ rates that could be compared to $H$, such as
the annihilation rate, which is slow, or the scattering rate, which is
fast. In the comoving rest frame, $2\to2$ processes redistribute
energy because they do not take place in the centre-of-mass
frame. They can bring a group of particles into kinetic (but not
chemical) equilibrium.

The annihilation cross section  for $\chi$s of energy $ \simeq 
m_\phi/2$ is
\begin{equation}
 \sigma_{\ann}  \simeq \frac{16 N_{\c} \alpha^2}{m_\phi^2},
\end{equation}
while
the cross-section for scattering (see figure \ref{f2}) is infrared
divergent:
\begin{equation}
 \sigma_{\scat} \simeq \frac{ N_{\c}  \pi \alpha^2}{m_\phi^2}
\int \frac{\rmd (\sin \theta)}{(1 - \cos \theta)^2}   = 2 \pi N_{\c} \alpha^2
 \int  \frac{ \rmd { t}}{{ t}^2}
\label{sigma}
\end{equation}
where $\theta$ is the scattering angle between $k$ and $k'$ and $t$ is
the 4-momentum transfer squared. Note that there is no thermal bath
whose plasma frequency can provide a cutoff for the scattering
cross-section (we are trying to compute when the thermal bath
appears). In the following section, we will show that the
thermalisation rate, due to scattering, is the rate 
 $\sim\,n_{\chi} \sigma_\ann$ associated with
annihilations.

The estimates in this section assume that
$\alpha_{\phi}\lesssim\,m_{\phi}/M_{\Pl}$, and
$m_{\phi}\lesssim\,M_{\GUT}$. The assumption that inflaton decay
products need to lose energy and produce particles to reach a thermal
distribution breaks down at $\alpha_{\phi} \sim\,m_{\phi}/M_{\Pl}$. If
we approximate the equilibrium number density $n_{\eq} \sim\,T^3 \sim
\rho^{3/4}$, we find, at $T_{\reh}$:
\begin{equation} 
 \frac{n_{\chi}}{n_{\eq}}
 \simeq \frac{\rho_{\phi}^{1/4}}{m_{\phi}} \simeq
 \sqrt{\frac{\alpha_{\phi} M_{\Pl}}{m_{\phi}}} \simeq
 \frac{T_{\reh}}{m_{\phi}},
\label{Tm} 
\end{equation} 
so $n_{\chi} \ll \, n_{\eq}$ if
$\alpha_{\phi} \ll  \, m_{\phi}/M_{\Pl}$. Perturbative particle
interaction rate estimates suggest that gauge interactions are not
fast enough to be in equilibrium at energy densities of ${\cal
O}(10^{15}\GeV)^4$, so our discussion breaks down at these energies.

\section{energy loss in elastic scattering}
\label{sect3}

From the perspective of a $\chi$ particle produced in the decay of a
$\phi$, thermalisation is a process of losing energy. So the
thermalisation timescale for $\chi$ is the timescale over which it can
lose energy $\simeq m_\phi-T_{\reh}\sim\,m_\phi$ to the surrounding
particles. We can estimate this timescale by integrating the
particle's rate of energy loss via scattering. In this section, we
consider elastic scattering.  We neglect annihilations, which we do
not expect to qualitatively affect our calculation, because hard
scattering processes are included. Scattering interactions have been
neglected in many previous thermalisation papers, who take the $2\to2$
annihilation rate as the thermalisation rate. We show here that the
thermalisation rate due to scattering is logarithmically enhanced over
the annihilation rate. So taking the $2\to2$ thermalisation rate
$\Gamma_{\therm}\sim\Gamma_{\ann}$ is justified, but claims that
scattering is irrelevant for thermalisation are not. We neglect the
cosmological expansion for the following two sections to avoid the
scale factor cluttering up formulae.

The elastic scattering cross-section (\ref{sigma}) diverges as $
\sigma\propto\rmd\theta/\theta^3$, where $\theta$ is the scattering
angle. The energy exchanged is $\Delta E \simeq m_\phi \theta^2$/2, so
the energy loss rate has a softer infrared divergence. The
cross-section for soft glancing scattering may be very large, but the
energy exchange is small, which reduces the importance of soft
processes for thermalisation.

The rate of energy loss of a particle scattering on a thermal bath
\begin{equation}
 \frac{{\rmd}E}{d \tau} = \langle n \sigma_{\scat} \Delta E \rangle ,
\label{dE}
\end{equation}
where $n\sim\,T^3$ is an equilibrium number density and $\Delta\,E$ is
the energy exchanged, has been extensively studied in finite
temperature field theory \cite{bj,FTFT}. The logarithmically
divergent result is
\begin{equation} 
 \frac{ {\rmd}E}{d \tau} 
 \sim 2 \alpha^2 T^2 \log \left(\frac{q_{\max}}{q_{\min}} \right) ,
\label{bj}
\end{equation}
where $q_{\max}$ and $q_{\min}$ are the maximum and minimum momentum
transfer respectively.

We follow ref.\cite{bj} in making a rough estimate of
${\rmd}E/{\rmd}\tau$ for a $\chi$ scattering on an unthermalised bath
of inflaton decay products. Consider the kinematics of scattering in
any frame where $\vec{p}\cdot\vec{q}\simeq0$ (or the approximation
where $\vec{p}\cdot\vec{k}'=(k_0'/k_0)\cdot\vec{p}\cdot\vec{k}$);
see figure 2 for the momenta of the particles involved. We are
interested in the energy lost by the relativistic $\chi$ with incident
momentum $k$. If the 4-momentum exchanged is $q^2=(k-k')^2=-t$,
\footnote{$s,t$ and $u$ are the kinematical variables, $\tau$ is
time.} then 

\begin{equation} 
\frac{q_0}{k_0}=\left\{
\begin{array}{ll} 
\frac{2 p \cdot q}{{ s}} (1 - \cos \varphi) &, {\rm ~for~}
                            \vec{p} \cdot \vec{q} = 0 \\
\frac{2 p \cdot q}{{ s}} &, {\rm ~for~}
\vec{p} \cdot \vec{k}' = 
\frac{k_0'}{k_0} \vec{p} \cdot \vec{k} ~~,
\end{array}
\right.
\label{lorentz}
\end{equation}
where $\varphi$ is the angle between the incident particles, and
$s=2k_0\,p_0(1-\cos\varphi)$. For $p'$ on-shell,
$q_0\simeq\,k_{0}t/s.$ In the relativistic limit where we can neglect
particle masses, equation (\ref{dE}) can be written as \cite{bj}
\begin{equation}
\frac{1}{k_0} \frac{dk_0}{d \tau} \simeq
 \int \frac{\rmd^3 p}{(2\pi)^3} f(p)
 (1 - \cos\varphi) \int \rmd{t} \frac{2\pi N_{\c}\alpha^2}{t^2} 
 \frac{(k_0 - k_0')}{k_0} ,
\label{dEdt1}
\end{equation}
where we have kept only the most infrared divergent
contributions. Here $f(p)$ is the momentum space distribution of the
$\chi$s, and $\rmd\sigma=2\pi\,N_{\c}\alpha^2 {\rmd}t/t^2$ is the
scattering cross-section. Using $t/s=(k_0-k_0')/k_0$, one finds
\cite{bj}
\begin{equation}
 \Gamma_\elas = \frac{1}{k_0}\frac{\rmd k_0}{\rmd \tau} \simeq 
 \frac{16 \pi N_{\c} \alpha^2 n_{\chi}}{m_\phi^2} 
 \ln\frac{m_\phi}{T_{\reh}}
 \sim \frac{2\pi N_{\c}\alpha^2 \alpha_{\phi}^2 M_{\Pl}^2}{m_\phi} 
 \ln\frac{m_\phi}{T_{\reh}} .
\label{dEdt}
\end{equation}
We took $q_{\min}\sim\,T_{\reh}$; we discuss why in the next
section. The last equality is $\Gamma_\elas$ evaluated at
$T_\reh$. This is logarithmically enhanced with respect to the 
hard annihilation rate $\Gamma_{\ann}\sim\alpha^2n_{\chi}/m_{\phi}^2$ . 

It is easy to see in equation~(\ref{dEdt}) that the infrared
divergence of the scattering cross-section
$\sim\alpha^2\int\rmd{t}/t^2$ is removed from the energy exchange rate
because $q_0\propto\,t=-q^2$ (rather than
$q_0\propto\sqrt{t}$). Elastic scattering is an inefficient energy
exchange mechanism, so only speeds up thermalisation by a logarithmic
factor with respect to hard annihilation processes. This agrees with
the numerical results of ref.\cite{enq}, who counted the number of
scatterings required to bring their bath of particles to thermal
equilibrium. They found that a few hard collisions per particle were
sufficient, but that more soft interactions were required.

We  estimate the timescale for a $\chi$ particle to lose an energy
of ${\cal O}(m_\phi)$ by elastic scattering with other $\chi$s to be
\begin{equation}
 \tau_{\elas} \sim \left(\frac{2}{m_\phi}\frac{\rmd k_0}{\rmd\tau}
 \right)^{-1}
 \simeq \left[\frac{32\pi N_{\c}\alpha^2}{m_\phi^2} n_\chi
 \ln\left(\frac{m_\phi}{T_{\reh}} \right)\right]^{-1}.
\label{tau1}
\end{equation}

\section{$2\to3$ scattering}
\label{sect4}

There are at least two flaws in the estimate leading to equation
(\ref{tau1}). Firstly, new $\chi$ particles must be created to bring
the $\chi$s into chemical equilibrium. The average comoving energy per
$\chi$ particle will remain $m_\phi$ until it can be redistributed
among newly created particles. So the rate of energy exchange, which
we calculated, will only be the rate of energy loss for an average
$\chi$ if there are newly created $\chi$ particles available to absorb
the energy. Secondly, since elastic scattering is an inefficient
energy transfer process, inelastic processes, although higher order in
$\alpha$, could be faster.

There are various $2\to3$ interactions by which $\chi$s can lose
energy and produce particles. We do not worry about the spin of the
particles created, because this section is based on dimensional
analysis. We identify particles by their energy---$\chi$s are inflaton
decay products, and $g$s are the particles of energy $\sim\,T_\reh$
being created to populate the thermal bath. We refer to $g$s as gauge
bosons, although there will be $\chi$ particles of energy $T_\reh$ in
the thermal bath. We make a separation based on energy because
cross-sections and number densities depend on it. We neglect gauge
bosons with energy less than $T_{\reh}$ because they would
subsequently need to be scattered up in energy to attain the
equilibrium relation $\langle{E}\rangle\simeq 3T$. The $\chi$
particles can produce gauge bosons and lose energy via
$\chi\bar{\chi}\rightarrow\chi\bar{\chi}g$,
$\chi\chi\rightarrow\chi\chi\,g$ and
$\chi\,g\rightarrow\chi\,g\,g$. The first process is $s$-channel, with
$s \sim\,m_\phi^2$, so we concentrate on the last two $t$-channel
processes.

There are many Feynman diagrams contributing to
$\chi\chi\rightarrow\chi\chi\,g$, which can be found, with the
associated matrix elements, in ref.\cite{J+R}. We estimate the
cross-section from the diagram of figure 2 with an outgoing gauge
boson of momentum $\ell\,(\ell^2=0$) emitted from the leg $p'$. So
$p^{'2}=(p''+\ell)^2=W^2$ is off-shell, and $p^{''2}=m_{\chi}^2$. We
neglect gamma matrices in the matrix element squared, and estimate
\begin{equation}
 \sigma_{\chi \chi \rightarrow \chi \chi g} 
 \sim 2 \pi N_{\c} \alpha^3 \int \frac{\rmd{t}}{t^2}\frac{{\rmd}W^2}{W^2}
 \sim \frac{\alpha^3}{T_{\reh}^2} \log\left( 
 \frac{m_\phi^2}{T_{\reh}^2} \right) ,
\label{sigma23}
\end{equation} 
where $(k-k')^2=-t,1/t^2$ corresponds to the photon propagator, and
$1/W^2$ to the fermion propagator. We treat $W^2$ and $t$ as
independent variables, ranging from $ T_{\reh}^2$ to $m_\phi^2$.  We
take the lower limit to be $T_{\reh}^2$ because we are interested in
making $n_g\sim\,T_{\reh}^3$ gauge bosons of energy
$\sim\,T_{\reh}$. Equation~(\ref{sigma23}) might capture the physics
of the infrared behaviour we are interested in. As $t\rightarrow0$, it
has the logarithmic divergence one expects from bremmstrahlung and a
quadratic divergence due to $t$-channel exchange. The remainder of
this paper relies on equation (\ref{sigma23}) being a reasonable
approximation; our arguments will not be true if
$\sigma_{\chi\chi\rightarrow\chi\chi\,g}$ is suppressed by a additional
factors of $T_{\reh}/m_\phi$ due to phase space or cancellations in
the matrix element.

One gauge boson for every $\chi$ can be generated via $\chi\chi$ 
scattering in a time 
\begin{equation}
 \tau_\inel \sim 
 \left(n_{\chi} \sigma_{\chi \chi \rightarrow \chi \chi g}\right)^{-1} 
 \sim \left[\frac{ \alpha^3 T_{\reh}^2}{m_{\phi}} \right]^{-1}
 \sim \left(\frac{ T_{\reh}^2}{ \alpha m_{\phi}^2 }\right) \tau_\elas .
\label{tauprod}
\end{equation}
To make $n_g\sim\,T_{\reh}^3$ gauge bosons, we need to make
$m_\phi/T_{\reh}$ gauge bosons for each $\chi$ particle, as can be
seen from equation~(\ref{Tm}). So the timescale to produce a 
thermal distribution of gauge bosons  via $\chi \chi$ scattering can 
be estimated as $\left(\frac{m_{\phi}}{T_\reh} \right)
\tau_\inel$. However, equation
(\ref{tauprod}) is also the timescale for the
particles $g$ to thermalise among themselves, at a rate
$\sim\alpha^3 n_{\chi}/T_\reh^2$. Once $n_g>n_{\chi}$,
gauge bosons can be produced in $ \chi g \rightarrow \chi g g$
and $n_g$ will grow rapidly, as observed
in reference \cite{enq}. We can write
\begin{equation}
\frac{{\rmd}n_g}{\rmd \tau} \sim \sigma_{\chi\chi\rightarrow\chi\chi g}
 ( n_{\chi}^2 + n_{\chi} n_g) .
\label{2star}
\end{equation}
The first term will dominate until $n_g\sim\,n_{\chi}$ at $\tau_\inel$, 
then $n_g$ will grow exponentially
with a timescale $\tau_{\inel}$. So the timescale to produce a thermal
number density using the cross-section (\ref{sigma23}) is
equation (\ref{tauprod}).

Cooling via the $2\to3$ scattering cross-section (\ref{sigma23}) will
be faster than the $2\to2$ cooling rate computed in the previous
section. We estimate the energy loss of the $\chi$ which emits a gauge
boson to be of order $T_{\reh}\sim\sqrt{t}$, so
\begin{equation}
\frac{{\rmd}E}{\rmd\tau} 
 \sim\,\vev{(n_{\chi} + n_g) \sigma_{\chi \chi \rightarrow \chi \chi g} 
   T_\reh}
 \sim \frac{ \alpha^3 (n_{\chi} + n_g)}{ T_{\reh}}  .
\label{approx}
\end{equation}
The timescale to lose energy $\sim\,m_{\phi}$ through scattering on
$\chi$s will be $\tau \sim $ $ [\alpha^3 n_{\chi}/(m_{\phi}
T_\reh)]^{-1}$. This is already a factor of $T_\reh/(\alpha m_{\phi})
$ shorter than the elastic cooling timescale (\ref{tau1}). However,
the $\chi$s can lose energy even faster by scattering off the growing
bath of gauge bosons.  If we substitute $n_g \sim\,n_{\chi}(a_\reh)
{\rme}^{\tau/ \tau_{\inel}}$ (approximately the solution of
equation~(\ref{2star})) into equation~(\ref{approx}), we find that the
timescale for a $\chi$ to lose energy $ \sim\,m_{\phi}$ is 
$\tau\sim\tau_\inel$. This is the timescale (\ref{tauprod}) to produce a
thermal bath, and equals $T_\reh^2/(\alpha m_{\phi}^2)\times\tau_\elas$.
We expect $\alpha m_\phi^2/T_{\reh}^2>1$, so $2\to3$ interactions cool 
the $\chi$s faster than elastic scattering, despite being higher order 
in $\alpha$.

There are two disturbing features to our estimate of the thermalisation rate
$\Gamma_\inel = \tau_\inel^{-1}$. It is larger than the 
elastic rate, which is lower
order in $\alpha$, and it is infra-red divergent. We can estimate the
cooling timescale due to $2\to3$ processes of an energetic particle
incident on a thermal bath. In this case, $\rmd\,E/\rmd\tau$ due to
$2\to3$ scattering is $\sim\alpha^3 T^2$, an ${\cal O}(\alpha)$
correction to equation~(\ref{bj}). This is reassuring, because this
thermal energy loss rate has been carefully studied \cite{bj,FTFT}. We
find a faster thermalisation rate for our inflaton decay products at
${\cal O} (\alpha^3)$ than at ${\cal O} (\alpha^2)$ because the energy
exchanged and target number densities are larger in the inelastic
case.  Our inelastic rate is one factor of $m_{\phi}/T_\reh$ larger
than the elastic rate because $n_g > n_{\chi}$; the $\chi$ scatters
more frequently off the bath of particles it created in earlier
interactions. It is a second factor of $m_{\phi}/T_\reh$ larger
because more energy is transfered in an inelastic collision than in an
elastic one. We have two small parameters in our reheating problem:
$\alpha$ and $T_\reh/m_{\phi}$. The inelastic thermalisation rate is
higher order in $\alpha$ than the elastic rate, but lower order in
$T_\reh/m_{\phi}$.

Equation~(\ref{approx}) has an infra-red divergence: the
rate for cooling by emission of gauge bosons of energy $\mu$ diverges
as $1/\mu^2$. Physical observables should not be infrared divergent, so
a cutoff is required for our estimate. There is initially no thermal bath
present  to justify using the ``thermal mass $\sim
gT$''---but there will be at the end of (re)heating, so we imagine
that the final state gauge boson must have energy of order $T_{\reh}$
to be on-shell. To see this, suppose that in a time $\Delta\tau$ a
fraction $\mu^3/n_{\chi}$ of the $\chi$s scatter inelastically, emitting 
a gauge boson of energy $\mu$. This creates a bath of gauge bosons with
number density $n_g(\mu)\sim\mu^3$, through which the next 
generation gauge bosons  must propagate. So these next
generation gauge bosons must have
energy $\gtrsim\mu$. As the time interval $\Delta\tau$ lengthens,
the cutoff $\mu$ grows to $T_{\reh}$. We therefore require the final
state gauge bosons to have energy $\gtrsim T_{\reh}$, and assume that
the momentum transfer in the scattering is of order the energy of the
emitted gauge boson. The exchanged gauge boson may not
have a ``thermal mass'' due to interactions with other gauge bosons,
because it may not live long enough to interact with them.

\section{Thermalisation}
\label{sect5}

We would like to know the thermalisation timescale $ \tau_{\therm}$
after reheating. We estimate $\tau_{\therm} = \Gamma_{\therm}^{-1}$,
and identify the thermalisation rate as the rate of energy loss for a
$\chi$, or the rate of particle production. (The two turn out to be
comparable). In this section, we estimate when the Universe will be
thermal, using the rates from the previous two sections.  In the first
part, we discuss two-to-to rates \cite{EENO,scott,McT,R}. We review the
thermalisation mechanism suggested in reference 
\cite{McT}, and explain where we
disagree with those thermalisation estimates. We make some remarks on
using decay processes as a particle production mechanism.  In the
second part, we compute the upper bound on the inflaton mass, below
which the Universe will be thermalised at $T_\reh$ due to $2\to3$
scattering processes.

\subsection{$\chi \chi \rightarrow \chi \chi$}

The elastic rate (\ref{dEdt}) (without the log term , which we also drop) has
been taken as the thermalisation rate in some previous work
\cite{EENO,R}. This assumes that a thermal number density is rapidly
produced by the decays of particles involved in the $2\to2$
processes. The rate scales after $T_\reh$ as $a_\reh/a$. We start in
the instantaneous decay approximation to review thermalisation
estimates based on equation~(\ref{dEdt}) \cite{EENO,McT,R}.
Thermalisation will happen immediately at $T_{\reh}$ if
\begin{equation} 
\alpha_{\phi} > \frac{ m_{\phi}^2}{4 N_{\c} \alpha^2
M_{\Pl}^2} . 
\label{abd} 
\end{equation} For $m_{\phi}$ of ${\cal
O}(10^3) $ GeV, the bound (\ref{abd}) translates into
$\alpha_{\phi}>10^{-30}$ which is certainly satisfied. However, for
$m_{\phi}\sim\,M_{\GUT}\sim10^{16}\GeV$, $\alpha_{\phi}>10^{-4}$. For
an inflaton mass of order the hidden sector scale $10^{12}\GeV$,
equation~(\ref{abd}) implies that the inflaton decay products can
annihilate with each other within a Hubble time if
$\alpha_{\phi}>10^{-12}$, which may not necessarily be the case
\cite{EENO,R}. For instance, if $\phi$ decays gravitationally,
$\alpha_{\phi}\sim\,m_{\phi}^2/M_{\Pl}^2\sim10^{-14}$. If the
$\phi\chi\bar{\chi}$ interaction is of electron yukawa strength, then
$\alpha_{\phi}\simeq10^{-14}$.

Equation~(\ref{abd}) appears peculiar, because it gives a {\it lower}
bound on $\alpha_{\phi}$. At smaller $\alpha_{\phi}$, the inflaton
decays later, so the number density of inflatons is redshifted and
therefore the number density of $\chi$s is similarly smaller
($n_{\chi} \simeq n_{\phi})$. The energy of the $\chi$s remains
$m_{\phi}/2$, so the $\chi$ interaction rate is smaller (scales as
$\alpha_{\phi}^2$, see equation (\ref{dEdt})). The expansion rate $H$ is also
smaller when $\alpha_{\phi}$ is smaller, but this is a less important
effect because $H \simeq \alpha_{\phi} m_{\phi}$ at $T_{\reh}$.

We now relax the instantaneous decay approximation and consider
whether the $\chi$s produced between $T_{\max}$ and $T_{\reh}$ have
time to annihilate. Repeating the estimate that lead to
equation~(\ref{dEdt}), using $n_{\chi}$ from equation~(\ref{nchi})
with $(1-{\rme}^{-\Gamma\tau})\simeq\Gamma\tau$, and taking
$E_{\chi}=m_{\phi}/2$ gives 
\begin{equation} 
 \Gamma_\elas(a) \simeq
 \frac{32 N_{\c} \alpha^2}{m_{\phi}^2} \frac{3 H^2(a) M_{\Pl}^2}{8 \pi
 m_{\phi}} \frac{2 \Gamma_{\phi}}{3 H(a)}.
\label{Gmax}
\end{equation}
If equation (\ref{abd}) is satisfied, $\Gamma_\elas>H$ from $T_{\max}$
onwards; that is, if $\Gamma_\elas > H$ is true at $T_{\max}$ if it is
true at $T_{\reh}$, and vice-versa. $\Gamma_\elas/H$ is a constant
between $T_{\max}$ and $T_{\reh}$ because $\Gamma_\elas$ scales as
$n_{\chi} \sim a^{-3/2}$, and $H^2$ scales as $\rho_{\phi} \sim
a^{-3}$.

Now let us suppose that equation (\ref{abd}) is {\it not} satisfied. As
discussed in refernce 
 \cite{McT}, the annihilation rate among the first $\chi$s
produced at $T_{\max}$ grows relative to the expansion rate $H$. So it
is claimed \cite{McT} that these particles can interact and
produce particles at some intermediate scale factor $a_{\max} < a <
a_{\reh}$.  More energetic particles can then thermalise rapidly, when
they are produced, by interacting with this soft tail after it has
reached kinetic and chemical equilibrium. This scenario is dubbed
catalysed thermalisation \cite{McT}. However we disagree with some of the
estimates in reference \cite{McT}, because the annihilation rate of a $\chi$ of
energy $E_2$ with less energetic $\chi$s is always smaller than the
annihilation rate with more energetic $\chi$s.  We can see this by
evaluating those two rates at $T_{\reh}$: 
\begin{equation}
  \int_0^{E_2} {\rmd}E 
 \frac{ {\rmd}n}{{\rmd}E} \frac{\alpha^2}{E E_2} 
 < \int_{E_2}^{m_{\phi}} {\rmd}E \frac{ {\rmd}n}{{\rmd}E} 
 \frac{\alpha^2}{E E_2}
 \sim 48 N_{\c} \frac{\alpha^2}{ m_{\phi} E_2}n_{\chi} (a_{\reh}).
\label{no}
\end{equation}
We agree that the less energetic $\chi$s are more likely to interact
(the right hand side of equation~(\ref{no}) is larger than that of
equation~(\ref{dEdt})), but they will annihilate with one of the
energetic $\chi$s whose density is higher, hence their energy will
increase. It will not substantially cool the energetic $\chi$s,
because few of them interact with a less energetic $\chi$. Nonetheless
it is possible that a thermalised seed is formed out of the less
energetic $\chi$s. Thermalisation is a process of energy loss and
particle production, which are both assumed to occur rapidly due to
decays in the annihilation and decay scenario of \cite{McT,R}.
Suppose that $\chi$ cannot decay, but can annihilate into some rapidly
decaying particle $\rho$. An efficient cascade decay of $\rho$ could
then produce many lower energy particles that rapidly thermalise. The
mass of at least one of the particles at every step of the cascade
must be heavy enough that their decay rate
$\Gamma_{\chi}=\lambda^2m_{\chi}^2/E_{\chi}$ is larger than
$H$. This requires some tuning of the mass, and a sufficiently
large $\lambda$, but is certainly possible, particularily in
supersymmetric theories where flat direction vevs could provide such
an intermediate scale mass.  We do not  discuss further the scenario of
thermalisation via annihilations and decays \cite{McT,R}. In the next
subsection, we focus on the generation of particles by $ 2\to3$ gauge
scattering interactions, which should occur generically  in all models.

\subsection{$2\to3$ interactions}

In section~\ref{sect4}, we estimated the energy lost by a $\chi$
scattering and emitting a gauge boson to be $\Gamma_\inel \sim\alpha
(m_{\phi}/T_\reh)^2 \Gamma_\elas$.  One factor of $ (m_{\phi}/T_\reh)$
arises because we took the inelastic energy loss in a collision to be
$T_\reh$, rather than $T_\reh^2/m_{\phi}$ as in the elastic case. The
second factor of $m_{\phi}/T_\reh$ comes from scattering the $\chi$s
off the denser thermal bath being produced in the $2\to3$
interactions. We expect that $\alpha \sim 1/30$, and that
thermalisation would take place within a Hubble time if $T_{\reh} \sim
m_{\phi}/6$. So $\alpha\,m_{\phi}^2/T_{\reh}^2>1$, energy loss via $2
\rightarrow 3$ processes is faster than via $2\to2$ elastic
scattering, and we estimate the thermalisation rate at $T_{\reh}$ to
be $\Gamma_{\therm}=\Gamma_{\inel}=\tau_\inel^{-1}$ (see equation
\ref{tauprod}). This coincides with the rate to create a number
density $n_g\sim\,T_{\reh}^3$ of gauge bosons via
$\chi\chi\rightarrow\chi\chi\,g$ and
$\chi\,g\rightarrow\chi\,g\,g$. Thus $\Gamma_{\inel}$ will exceed
the expansion rate  at $T_\reh$ if
\begin{equation}
 m_{\phi} \lesssim \alpha^3 {M_{\Pl}} .
\end{equation}
The COBE results require the inflationary scale
(and hence $m_{\phi}$) to be much less than $10^{16}$ GeV \cite{lss}, so 
this condition ought to be satisfied. As expected, this is 
a weaker bound than equation~(\ref{abd}). It is independent 
of $\alpha_{\phi}$ because both $\Gamma_\inel(a_\reh) $ 
and $H(a_\reh)$ are proportional to  $\alpha_{\phi}$.

Note that $2\to3$ interactions can thermalise the relativistic
particles at $T_\max$ within a Hubble time if
$\alpha_{\phi}>m_{\phi}^2/(\alpha^3 M_{\Pl}^2)$, which is similar to
the condition (\ref{abd}).

The rate of energy loss of an energetic particle incident on a thermal
bath is ${\rmd}E/dt \sim\alpha^2 T^2$ (see equation \ref{bj})
\cite{bj,FTFT}. The thermalisation timescale for a particle of energy
$m_{\phi}$ is therefore $\sim (\alpha^2 T^2/m_{\phi})^{-1}$. Our
estimate of $2\to3$ interaction rates suggests that reheating a cold
Universe is a factor of $\alpha$ slower; the inflaton decay products
rapidly produce a bath of soft particles, and cool by interacting with
them.

\section{conclusion}
\label{sect6}

If the inflaton $ \phi$ decays perturbatively in a cold Universe, its
decay products must interact to thermalise. There are two aspects to
thermalisation: producing additional particles and distributing energy
among them, so as to obtain a kinetic and chemical equilibrium
distribution in momentum space. It is often assumed that
thermalisation proceeds by annihilations and decays, so the
thermalisation timescale is taken to be the timescale of hard
annihilations among inflaton decay products. In this paper we
discussed thermalisation via scattering interactions. Our estimates
suggest that these soft processes lead to faster thermalisation.

We first considered energy loss via scattering between inflaton decay
products $\chi$, and found that the timescale for a $\chi$ to cool
down to $T_\reh$ is the timescale of hard annihilations. The
scattering cross-section is infrared divergent so the interaction rate
is large, but little energy is exchanged in soft elastic scattering so
it does not lead to rapid thermalisation.

We then estimated the cooling rate via inelastic $2\to3$ processes
$\Gamma_\inel$, and found it to be much larger: $ \Gamma_\inel \sim
\alpha m_{\phi}^2/T_\reh^2 \times$ the elastic rate. We also estimated
the timescale to create a particle number density $\sim T_{\reh}^3$
via $2\to3$ interactions and found that it was of order $
\Gamma_\inel^{-1}$. Thus $2 \rightarrow 3$ 
 scattering is a more efficient thermalisation
process than $2\to2$ elastic scattering among inflaton decay products,
because the energy transfer in a collision is a factor of
$m_{\phi}/T_\reh$ larger, and because inflaton decay products can
collide with the particles they produced in earlier $2\to3$
interactions.

Our estimated thermalisation timescale  at $T_\reh$ is 
\begin{equation}
\tau_{\therm} 
\sim \left( \frac{\alpha^3 n_{\phi}}{ T_{\reh}^2} \right)^{-1}
\sim \left( \frac{\alpha^3 T_{\reh}^2}{m_\phi} \right)^{-1}
\end{equation}
where $n_\phi$ is the inflaton number density just before $T_{\reh}$.
This result is independent of the details of
the reheating model. If the Universe 
is dominated by relativistic particles of energy $E$,
then they will thermalise within a Hubble time
to a temperature $T_{\reh} \sim \rho^{1/4}$ if
\begin{equation}
\Gamma_\therm \sim \frac{\alpha^3}{\sqrt{\rho}} \frac{\rho}{ E} 
> H \Rightarrow   
E \lesssim \alpha^3 M_{\rm Pl} .
\end{equation}

\subsection*{Acknowledgements}
S.D. would like to thank Marta Losada, Tomislav Prokopec and 
Sharon Vadas for useful conversations.

\end{document}